\documentclass[prc,twocolumn,showpacs,floatfix,nofootinbib,preprintnumbers,superscriptaddress,amsmath,amssymb]{revtex4-1}

\usepackage{CJKutf8}
\usepackage{graphicx}
\usepackage{epsfig}
\usepackage{amsmath,amssymb}
\usepackage{bm}
\usepackage{color}
\usepackage{float}
\usepackage{dcolumn}

\begin{document}
\begin{CJK*}{UTF8}{gbsn}

\title{Kerman-Onishi conditions in self-consistent tilted-axis-cranking mean-field calculations}

\author{Yue Shi (石跃)}
\affiliation{Department of Physics and Astronomy, University of
Tennessee, Knoxville, Tennessee 37996, USA}
\affiliation{Joint Institute for Heavy Ion Research, Oak Ridge National Laboratory, Oak Ridge, Tennessee 37831, USA}

\author{C. L. Zhang (张春莉)}
\affiliation{Department of Physics and Astronomy, University of
Tennessee, Knoxville, Tennessee 37996, USA}
\affiliation{State Key Laboratory of
Nuclear Physics and Technology, School of Physics, Peking University, Beijing 100871, China}

\author{J. Dobaczewski}
\affiliation{Institute of Theoretical Physics, Faculty of Physics, University of Warsaw, ul. Ho{\.z}a 69, PL-00681 Warsaw, Poland}
\affiliation{Department of Physics, PO Box 35 (YFL), FI-40014
University of Jyv{\"a}skyl{\"a}, Finland}

\author{W. Nazarewicz}
\affiliation{Department of Physics and Astronomy, University of
Tennessee, Knoxville, Tennessee 37996, USA}
\affiliation{Physics Division, Oak Ridge National Laboratory, Oak Ridge, Tennessee 37831, USA}
\affiliation{Institute of Theoretical Physics, Faculty of Physics, University of Warsaw, ul. Ho{\.z}a 69, PL-00681 Warsaw, Poland}

\begin{abstract}
\begin{description}
\item[Background] For cranked mean-field calculations with arbitrarily oriented
rotational frequency vector $\boldsymbol{\omega}$ in the intrinsic frame, one has to employ
constraints on average values of the quadrupole-moment tensor, so as to keep the
nucleus in the principal-axis reference frame. Kerman and Onishi [Nucl. Phys. A {\bf
361}, 179 (1981)] have shown that the Lagrange multipliers that correspond to the
required constraints are proportional to $\boldsymbol{\omega} \times
\boldsymbol{J}$, where $\boldsymbol{J}$ is the average angular momentum vector.

\item[Purpose] We study the validity and consequences  of the Kerman-Onishi
conditions  in the context of self-consistent tilted-axis-cranking (TAC) mean-field
calculations.

\item[Methods] We perform self-consistent two-dimensional-cranking
calculations (with and without pairing) utilizing the symmetry-unrestricted
solver {\sc hfodd}. At each tilting angle, we compare the calculated values of
quadrupole-moment-tensor Lagrange multipliers and $\boldsymbol{\omega} \times
\boldsymbol{J}$.

\item[Results] We show that in self-consistent calculations, the
Kerman-Onishi conditions are obeyed with high precision. Small
deviations seen in the calculations with pairing can be attributed to
the truncation of the quasiparticle spectrum. We also provide results
of systematic TAC calculations for triaxial strongly deformed bands
in $^{160}$Yb.

\item[Conclusions] For non-stationary TAC solutions, Kerman-Onishi
conditions link the non-zero values of the angle between
rotational-frequency and angular-momentum vectors to the constraints on
off-diagonal components of the quadrupole-moment tensor. To stabilize
the convergence of self-consistent iterations, such constraints have
to be taken into account. Only then one can determine the Routhian
surfaces as functions of the tilting angles.
\end{description}
\end{abstract}


\maketitle
\end{CJK*}

\section{Introduction}\label{intro}

Axially deformed nucleus can execute collective rotations with the angular
momentum built along the axis that is perpendicular to the symmetry
axis\,\cite{ring80,szy83}. Due to the larger moment of inertia, rotation about this
axis is energetically more efficient in generating the same amount of angular
momentum, compared to the mechanism of particle-hole excitations or
non-collective rotations\,\cite{voigt83}. The standard theoretical
framework to describe high-spin phenomena is the cranking
model, in which a cranking term, $-\omega_1
J_1$\,\cite{inglis54,ring80,szy83}, is added to the mean-field Hamiltonian.

In principle, a triaxial nucleus  can rotate about an axis that does not
coincide with the principal axis (PA) \cite{frau01,frau97}. The evidence for
such rotations has been very limited, however, mainly owing to very indirect
experimental information concerning triaxial shapes. Theoretical challenges
include (i) the fact that very few nuclei are predicted to be triaxial in their
ground states \cite{Mol08}, and predictions are very model dependent
\cite{Sny13}; (ii) triaxial minima in potential-energy surfaces  associated
with triaxial shapes are soft, and (iii) minima of Routhians in function of
the tilting angles are shallow~\cite{frisk87,Naz96b}, meaning the
rotational axis can easily change its direction. The latter two challenges
require the mean-field models to be able to handle various correlations in a
self-consistent manner.

In self-consistent  Hartree-Fock (HF) TAC calculations, when the rotational
axis moves away from the PA, the nucleus has to stay fixed in the PA system.
This can be realized by adding linear constraints that would guarantee  that
the resulting off-diagonal components of the inertia tensor vanish. In
Ref.\,\cite{kerm81}, Kerman and Onishi have shown that the corresponding
Lagrange multipliers depend on the angular momentum, rotational frequency,
and the quadrupole moments of the system through relation\,(3.6) in
Ref.\,\cite{kerm81}, referred to as the Kerman-Onishi (KO) conditions in the
following.
Subsequent applications seldom included such linear constraints, except for
those described in Refs.~\cite{kerm81,naza92,hori96,*oi98,*oi03}, in which
nuclear mean fields were either modeled by phenomenological potentials
or   the self-consistent fields of the pairing-plus-quadrupole Hamiltonian~\cite{hori96,*oi98,*oi03}.
In this paper we extend these studies to mean-field approaches based
on realistic energy density functionals (EDFs).

Recently, a multitude of high-spin bands in nuclei around $^{158}$Er have been
observed\,\cite{paul07,patt07,agui08,tea08,oll09,oll11,wang11} and interpreted
in terms of triaxial strongly deformed (TSD) structures predicted by the PA
cranking (PAC) calculations\,\cite{dude85, paul07,shi12,kar12,afan12}.
Specifically, in $^{158}$Er, two minima (dubbed TSD1 and TSD2) with similar
$\beta_2$ and $|\gamma|$ deformations, but opposite signs of $\gamma$, are
calculated to be separated by a pronounced barrier. In our previous
study~\cite{shi12}, we have demonstrated that by allowing the rotational axis
to tilt away from the PA, one of these minima becomes a saddle point, thus
clarifying the nature of the TSD1 and TSD2 bands. In our fully self-consistent
TAC Skyrme-HF (SHF) and HF-Bogoliubov (SHFB) calculations, we noticed that the
KO conditions, which in the past have never been practically implemented in
the EDF approaches, are numerically fulfilled with high precision. Here, we
provide an in-depth analysis of these conditions and discuss their physical
consequences.

The paper is organized as follows. In Sec.\,\ref{ko-cond}, we present a
re-derivation of the KO conditions and extend it to the framework of the
Kohn-Sham (KS) theory, where one does not employ Hamiltonian but an EDF.
Sec.\,\ref{model} explains the model used in the
present work and expresses the KO conditions in terms of conventions used by
the {\sc hfodd} solver. Section.~\ref{results} contains the results of
calculations. Finally, the conclusions of this work are given in
Sec.~\ref{conclusions}.

\section{Kerman-Onishi conditions}
\label{ko-cond}

The conditions for a general nuclear rotation around arbitrary axis were
proposed by Kerman and Onishi \cite{kerm81} within the time-dependent
variational method. In the following Subsection, to fix the notation,
we recall the original derivation that is based on the constrained HF
method.

\subsection{Kerman-Onishi conditions in the Hartree-Fock approach with a Hamiltonian}

Within the constrained HF method, the rotation is imposed by adding a
cranking term, $-\boldsymbol{\omega}\cdot \boldsymbol{\hat J}$ to the
Hamiltonian $\hat{H}$, where $\boldsymbol{\hat J}$ is the angular
momentum operator and $\boldsymbol{\omega}$ are three Lagrange
multipliers identified with the rotational frequency vector. The
intrinsic frame is defined by bringing the average values of the
symmetric second-rank tensor,
\begin{equation}\label{Qtens}
\hat{Q}_{ij}\equiv  x_i x_j ,
\end{equation}
to its PA.
This can be achieved by requiring that the average values of
three off-diagonal components of $\hat{Q}_{ij}$:
\begin{equation}
\hat{B}_k = \hat{Q}_{ij}, ~~~~~~~(ijk; \textrm{cyclic}),
\end{equation}
are zero in the HF state $|\Phi\rangle$:
\begin{equation}\label{Bcond}
\langle \Phi| \hat{\boldsymbol{B}}|\Phi\rangle=0.
\end{equation}

The resulting Routhian can be written as:
\begin{equation}
\hat{H}' = \hat{H} - \boldsymbol{\omega}\cdot {\boldsymbol{\hat J}}
-\boldsymbol{\lambda}\cdot \hat{\boldsymbol{B}},
\end{equation}
where $\boldsymbol{\lambda}$ are three Lagrange multipliers enforcing condition
(\ref{Bcond}), and the rotational frequencies $\boldsymbol{\omega}$
are determined from the angular-momentum condition:
\begin{equation}\label{Jcond}
\boldsymbol{J} = \langle
\Phi | {\boldsymbol{\hat J}} | \Phi \rangle.
\end{equation}
The original derivation of KO conditions is based on the fact that
for  the stationary HF state $|\Phi\rangle$ that minimizes the Routhian
\begin{equation}
\delta \langle \Phi| \hat{H}'|\Phi\rangle=0,
\end{equation}
the following condition holds:
\begin{equation}\label{commJ}
\langle \Phi| [\hat{H}',\boldsymbol{\hat J}]|\Phi\rangle=0.
\end{equation}

By assuming the rotational invariance of the Hamiltonian:
\begin{equation}\label{Hinv}
[\hat{H},\boldsymbol{\hat J}]=0,
\end{equation}
denoting the three diagonal components of (\ref{Qtens}) as
$\hat{D}_i \equiv \hat{Q}_{ii}$, and noticing that
\begin{equation}\label{BD}
\begin{aligned}
{[\hat{B}_i,\hat{J}_j]} & =  -i\hat{B}_k, \\
[\hat{B}_k,\hat{J}_k] & =  i(\hat{D}_i  - \hat{D}_j),
\end{aligned}
\qquad (ijk; \textrm{cyclic})
\end{equation}
one arrives at the original KO conditions:
\begin{equation}
\label{eqn3}
\lambda_k = \frac{(\boldsymbol{\omega} \times \boldsymbol{J})_k}{D_i - D_j} ~~~~~~~~ (ijk; \textrm{cyclic}),
\end{equation}
where $D_i = \langle
\Phi | \hat{D}_i | \Phi \rangle$.  Consequently,
nonzero values of
Lagrange multipliers $\boldsymbol{\lambda}$ imply that vectors
$\boldsymbol{\omega}$ and $\boldsymbol{J}$ are not parallel. We note
here that $\hat{B}_k$, $\hat{D}_i$, and $\lambda_k$ are numbered by
three Cartesian directions and thus we use for them the standard
bold-face notations ${\boldsymbol{\hat{B}}}$,
${\boldsymbol{\hat{D}}}$, and ${\boldsymbol{\lambda}}$, respectively;
however, under space rotations they do not transform as vectors.

\subsection{Kerman-Onishi conditions in a DFT approach with an energy density functional}

In the framework of
the KS approach of the DFT \cite{kohn65}, the basic entity is the energy density ${\cal H}(\boldsymbol{r})$, which gives the total binding energy of the system in terms of the one-body density matrix $\rho$:
\begin{equation}
E\{\rho\}=\int d^3\boldsymbol{r}{\cal H}(\boldsymbol{r})=E_k\{\rho\}+E_p\{\rho\},
\end{equation}
expressed in terms of kinetic and potential energy terms.
The minimization of $E\{\rho\}$ with respect to $\rho$ under conditions
(\ref{Bcond}) and (\ref{Jcond})
results in self-consistent KS equations:
\begin{equation}\label{KSEq}
[h'(\rho),\rho]=0,
\end{equation}
where
$h'=h - \boldsymbol{\omega}\cdot {\boldsymbol{\hat J}}
-\boldsymbol{\lambda}\cdot \hat{\boldsymbol{B}}$ and
the mean-field Hamiltonian $h$ is now given by the
functional derivative of $E\{\rho\}$:
\begin{equation}
h = \frac{\partial}
{\partial \rho}E\{\rho\} =t+\Gamma,
\end{equation}
with $t = \frac{\partial}
{\partial \rho}E_k\{\rho\}$  being the one-body kinetic term and  $\Gamma = \frac{\partial}
{\partial \rho}E_p\{\rho\}$ -- one-body potential-energy term.

It is important to realize that in the KS approach, neither many-body
Hamiltonian $\hat H$ nor wave function $|\Phi\rangle$ are well-defined entities, unless
the energy density ${\cal H}$ is explicitly derived from an effective  two-body
interaction \cite{Ben03}. Moreover, even if the actual approach is based on a
Hamiltonian, the self-consistent density $\rho$ and  the potential term
$\Gamma(\rho)$ obtained from Eq.~(\ref{KSEq}) usually break the original
symmetries of $\hat H$ (in particular  the rotational invariance) due to the
spontaneous symmetry breaking mechanism. Consequently, if $\hat H$  depends on
density, as in the case of  Skyrme and Gogny interactions, condition
(\ref{Hinv}) obviously does not hold.

Let us now consider the group SO(3) of rotations. The corresponding  transformation is represented by the unitary operator $\hat{R}(\boldsymbol{\theta}) = \exp(i\boldsymbol{\theta}\cdot\boldsymbol{\hat{J}})$, where
$\boldsymbol{\theta}$ is a three-dimensional rotation vector.
 As discussed in Refs.~\cite{Carlsson08,Roh10},
since the energy density is covariant with $\hat{R}(\boldsymbol{\theta})$, that is,
\begin{equation}
{\cal H}^R (\boldsymbol{r})= {\cal H}(\hat{R}^+\boldsymbol{r}\hat{R}),
\end{equation}
and the kinetic term is a scalar,
the potential energy is invariant
with respect to a unitary transformation of the density \cite{raim11}
\begin{equation}\label{Potinv}
E_p(\rho) = E_p(\hat{R}\rho \hat{R}^+).
\end{equation}
As a matter of fact, relation (\ref{Potinv}) applies not only to
the total potential energy but also to individual contributions to
$E_p$ associated with terms characterized by different coupling
constants.  Physically, Eq.~(\ref{Potinv}) simply means that the total
energy does not depend on the orientation of the intrinsic density in
space.

The first-order expansion in $\boldsymbol{\theta}$ yields \cite{raim11}
\begin{equation}\label{Eexp}
E_p(\hat{R}\rho \hat{R}^+) = E_p(\rho) + i\boldsymbol{\theta} \cdot {\rm Tr}(\Gamma [\boldsymbol{\hat J},\rho]).
\end{equation}
Consequently,
\begin{equation}\label{trj}
{\rm Tr}(\Gamma [\boldsymbol{\hat J},\rho])={\rm Tr}(\rho [\Gamma,\boldsymbol{\hat J}])
=\langle [\Gamma,\boldsymbol{\hat J}]\rangle =0,
\end{equation}
where the symbol $\langle \ldots \rangle$ means an average value
${\rm Tr}(\rho \ldots)$. A KS analog of the Hamiltonian expression (\ref{commJ})
can be written as
\begin{equation}\label{commJ1}
{\rm Tr}(\rho [h',\boldsymbol{\hat J}])={\rm Tr}(\boldsymbol{\hat J}[\rho, h'])=0,
\end{equation}
where we used the self-consistency condition (\ref{KSEq}). Finally, by combining
(\ref{trj}) and (\ref{commJ1}) we obtain
\begin{equation}\label{commJ2}
\langle {[} \boldsymbol{\omega}\cdot \boldsymbol{\hat J}
+ \boldsymbol{\lambda} \cdot \hat{\boldsymbol{B}} , {\boldsymbol{\hat J}} {]} \rangle=0,
\end{equation}
which, with the help of (\ref{BD}),  yields the  KO conditions
(\ref{eqn3}) in the KS case, with $\langle \Phi| \ldots |\Phi\rangle$
being replaced by ${\rm Tr}(\rho \ldots)$.

The extension to superfluid DFT, or self-consistent Hartree-Fock-Bogoliubov
(HFB) theory is straightforward: in the derivations above, the single-particle
density matrix $\rho$ is  replaced by the generalized 2$\times$2 density matrix
$\cal R$, also involving the abnormal (pair) density $\tilde\rho$
\cite{doba84}, and the mean-field $h$ is replaced by the 2$\times$2 HFB
Hamiltonian, containing the pairing field $\tilde h$. All other steps are
identical; one needs to remember that the one-body operators
$\hat{\boldsymbol{B}}$ and $\hat{\boldsymbol{J}}$ act only in the particle-hole
space, that is, the average values do not involve the abnormal density.

\section{The model}
\label{model}

In this work, all calculations were performed by using the
symmetry-unrestricted solver {\sc hfodd} (version 2.49t)\,\cite{schu12}. For
tests of KO conditions in $^{158}$Er and TAC calculations of TSD bands in
$^{160}$Yb, we performed the SHF calculation without pairing.
We used the Skyrme EDF SkM*\,\cite{bart82} and 1,000 deformed
harmonic-oscillator (HO) basis states with HO frequencies of $\hbar \Omega_x =
\hbar \Omega_y = 10.080$\,MeV (up to $N_x=N_y=15$ HO quanta) and
$\hbar \Omega_z = 7.418$\,MeV (up to $N_z=20$ HO quanta). Detailed
tests show that such a size of the basis provides us with a sufficient precision for the
quantities studied in this work.
Tests of the KO conditions in $^{158}$Er were
performed at rotational frequency of $\hbar\omega = 0.6$\,MeV and for
configuration denoted by $\nu[23,23,22,22] \otimes \pi[17,18,16,17]$.  The
configurations are  labeled by the numbers of states occupied in the four
parity-signature ($\pi, r$) blocks, in the convention of Ref.\,\cite{doba04}.

Technically, these numbers correspond to the conserved $x_2$-signature, and
thus are valid only for rotations around the $x_2$ axis. However, for a
two-dimensional cranking tilted within a symmetry plane of the matter distribution, the
$\boldsymbol{J}$-signature corresponding to the direction of the
angular-momentum vector remains a good quantum number for any tilting angle
$\theta$. Therefore, the given numbers of states can be understood as
pertaining to the $\boldsymbol{J}$-signature. Of course, for non-zero tilting
angles, the code must be run in the broken-$x_2$-signature mode, whereupon the
configurations are set by fixing numbers of states occupied in parity blocks
only.

To check the KO conditions in the presence of pairing, we also performed  TAC HFB
calculations for $^{110}$Mo, with Lipkin-Nogami pairing (HFB+LN) included\,\cite{doba04}. The neutron and proton
pairing strengths were chosen as $(V_0^{\nu}, V_0^{\pi}) = (-196,
-218)$\,MeV, with a cutoff energy in the quasiparticle spectrum
of $E_{\rm cut}=60$\,MeV. This choice of the specific nucleus and pairing
strengths was dictated by the fact that with this choice one obtains a well
defined triaxial solution with sizable pairing, which makes it
a very suitable  test case for the KO conditions in HFB.

To link Eq.\,(\ref{eqn3}) to total quadrupole moments of the nucleus,
we first rewrite it terms of spherical components of the quadrupole
tensor (\ref{Qtens}).
In the {\sc hfodd} convention for multipole moments, we have:
\begin{equation}
Q_{\lambda \mu}(\boldsymbol{r}) =
a_{\lambda \mu} r^{\lambda} Y^*_{\lambda \mu}(\theta, \phi),
\end{equation}
where $Y_{\lambda \mu}$ are the standard spherical harmonics in the convention
of Ref.~\cite{vars88} and normalization factors $a_{\lambda \mu}$
have been defined in Table 5 of Ref.~\cite{doba04}. Then we have explicitly,
\begin{subequations}
\begin{eqnarray}
\label{Q2}
Q_{20} &=& 2x_3^2 - x_1^2 -x_2^2             , \\
Q_{21} &=& x_3 x_1-ix_3 x_2                  , \\
Q_{22} &=& \sqrt{3}(x_1^2 - x_2^2 -2ix_1 x_2),
\end{eqnarray}
\end{subequations}
with $Q_{2-1}=-Q_{21}^*$ and $Q_{2-2}=Q_{22}^*$, and thus,
\begin{eqnarray}
\label{convention1}
x_1 x_2 &=& \frac{\sqrt{3}}{6} \Im {Q}_{2-2}, \\
\label{convention2}
x_1 x_3 &=& \Re {Q}_{21}, \\
\label{convention3}
x_2 x_3 &=& \Im {Q}_{2-1}, \\
\label{convention4}
x_1^2 - x_2^2 &=& \frac{\sqrt{3}}{3}\Re {Q}_{22}, \\
\label{convention5}
x_3^2 - x_2^2 &=& \frac{\sqrt{3}}{6} \Re {Q}_{22} + \frac{1}{2} {Q}_{20}, \\
\label{convention6}
x_3^2 - x_1^2 &=& -\frac{\sqrt{3}}{6} \Re {Q}_{22} + \frac{1}{2} {Q}_{20}.
\end{eqnarray}

By means of  Eqs.\,(\ref{convention1})-(\ref{convention6}), the KO conditions
(\ref{eqn3}) read:
\begin{subequations}
\begin{eqnarray}
-\lambda_1 x_2 x_3 &=& +\frac{\omega_2 j_3 - \omega_3 j_2}{\frac{\sqrt{3}}{6}\Re \langle {Q}_{22} \rangle + \frac{1}{2} \langle {Q}_{20} \rangle} \Im {Q}_{2-1} \nonumber \\
 &\equiv& -L'_{2-1} \Im {Q}_{2-1}, \\
-\lambda_2 x_1 x_3 &=& -\frac{\omega_3 j_1 - \omega_1 j_3}{-\frac{\sqrt{3}}{6}\Re \langle {Q}_{22} \rangle + \frac{1}{2} \langle {Q}_{20} \rangle} \Re {Q}_{21} \nonumber \\
 &\equiv& -L'_{21} \Re {Q}_{21}, \\
-\lambda_3 x_1 x_2 &=& -\frac{\omega_1 j_2 - \omega_2 j_1}{2\Re \langle {Q}_{22} \rangle} \Im {Q}_{2-2}  \nonumber \\
 &\equiv& -L'_{2-2} \Im {Q}_{2-2},
\label{eqn4}
\end{eqnarray}
\end{subequations}
which can be conveniently written in terms of  factors
$L'_{2-1}$, $L'_{21}$, and $L'_{2-2}$.

Following Ref.~\cite{shi12}, in the calculations presented in this study we only
allow the rotational axis to tilt from the $x_2$-axis into the
$x_1$-$x_2$ plane; that is, the cranking is two dimensional. In this
case, only one of the three constraints, $-L_{2-2}\Im {Q}_{2-2}$,
is active. The other two constraints, $\langle \Im {Q}_{2-1}
\rangle =0$ and $ \langle \Re {Q}_{21} \rangle =0$, are automatically
realized by enforcing the $x_3$-$T$-simplex symmetry~\cite{(Dob00a)}. Then, in each
iteration, $L_{2-2}$ is updated so as to guarantee that $\langle
{Q}_{2-2} \rangle = 0$. To obtain a precise value of this
constraint, we use the augmented Lagrange method~\cite{stas10}.
Upon convergence, one obtains values of $L_{2-2}$,
components of total angular momentum, $j_1$ and $j_2$,
and $\langle {Q}_{22} \rangle$. Inserting $j_1$, $j_2$,
and $\langle {Q}_{22} \rangle$ into
relation~(\ref{eqn4}), one obtains values of $L'_{2-2} = \frac{\omega_1 j_2 -
\omega_2 j_1}{2\Re \langle {Q}_{22} \rangle}$. By comparing $L_{2-2}$
and $L'_{2-2}$ one can assess the extent  the KO
conditions are fulfilled.

\section{results}
\label{results}
\subsection{Tests of the Kerman-Onishi conditions}
\begin{figure}[htb]
\includegraphics[width=0.45\textwidth]{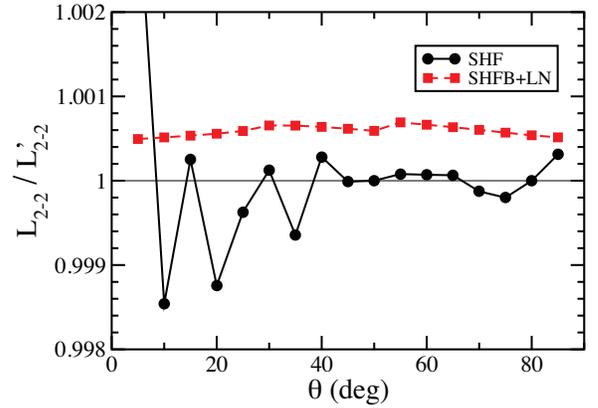} \caption{(Color
online) The $L_{2-2}/L'_{2-2}$ ratio for a TSD band in $^{158}$Er (SHE)
and a triaxial band in $^{110}$Mo (SHFB+LN) as a function of the tilting angle
$\theta$  defined in the  $x_1$-$x_2$ plane. The quantity $L_{2-2}$ is
 obtained self-consistently while  $L'_{2-2}$ is defined through
Eq.~(\protect\ref{eqn4}).}
\label{L-t}
\end{figure}

Figure~\ref{L-t} displays values of $L_{2-2}/L'_{2-2}$, calculated for the
triaxial  bands in $^{158}$Er and
$^{110}$Mo defined in  Sec.~\ref{model}, as a function of
the tilting angle $\theta$ in the $x_1$-$x_2$ plane. The quantity $L_{2-2}$ is obtained from
the self-consistent SHF (SHFB+LN) calculations, and $L'_{2-2}$ is obtained by means of
Eq.~(\ref{eqn4}). It can be seen that for the most of values of $\theta$, ratios
$L_{2-2}/L'_{2-2}$ stay inside the interval of [0.999-1.001], which means the
first 3 significant digits of $L_{2-2}$ and $L'_{2-2}$ are the same. We can thus see that in our  HF and HFB calculations the KO conditions are
fulfilled with a rather high precision. We note that at $\theta=0$ or 90$^{\circ}$, the
solutions correspond  to the uniform PA rotation, and thus the values of both
$L_{2-2}$ and $L'_{2-2}$ tend to zero; hence, the ratios $L_{2-2}/L'_{2-2}$ cannot be determined.

\begin{figure}[htb]
\includegraphics[width=0.45\textwidth]{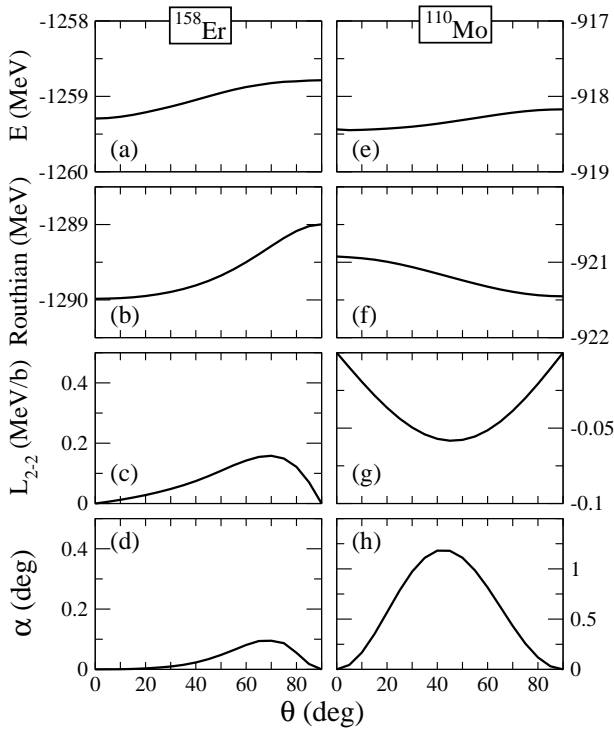}
\caption{From top to bottom: energies, Routhians, $L_{2-2}$, and
$\alpha$ (angle between $\boldsymbol{\omega}$ and $\boldsymbol{J}$)
as functions of the tilting angle $\theta$ for triaxial bands in  $^{158}$Er (left)
and $^{110}$Mo (right).}
\label{E-R-L-a}
\end{figure}
The energetics governing TAC rotations is illustrated in  Fig.~\ref{E-R-L-a}, which shows  energies, Routhians, $L_{2-2}$, and the angle $\alpha$ between $\boldsymbol{\omega}$ and $\boldsymbol{J}$
 as functions of  $\theta$ for $^{158}$Er and
$^{110}$Mo. It is seen that in these two triaxially deformed systems, stationary
solutions  appear only at $\theta=0^{\circ}$ or $90^{\circ}$.  The value of
$L_{2-2}$ vanishes at stationary points,  and it becomes finite
 when the Routhian has a nonzero slope as a function of $\theta$.
The value of angle $\alpha$ always remains fairly small, in $^{158}$Er and
$^{110}$Mo reaching 0.1 and 1$^{\circ}$, respectively.

\begin{figure}[tb]
\includegraphics[width=0.45\textwidth]{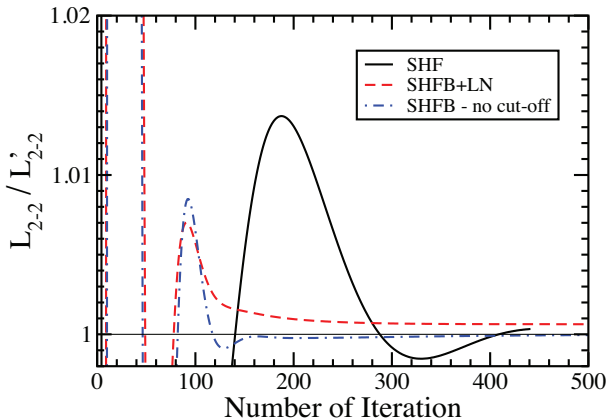}
\caption{(Color online) The ratio $L_{2-2}/L'_{2-2}$  as a function of
iteration number at $\theta=40^{\circ}$ for SHF, SHFB+LN, and SHFB (no cut-off).  See text for details.}
\label{L-i}
\end{figure}

As seen in Fig.~\ref{L-t}, there appear small residual deviations between
$L_{2-2}$ and $L'_{2-2}$ that need to be understood.  To this end, in
Fig.~\ref{L-i} we show the ratio $L_{2-2}/L'_{2-2}$ calculated at
$\theta=40^{\circ}$ as a function of the number of iterations. We see that in
the SHF case the convergence of the ratio  to the
exact value of $L_{2-2}/L'_{2-2}=1$  can be slow. Because of that, by stopping the iteration
when the convergence at every $\theta$ reaches a fixed stability, one obtains
values of $L_{2-2}/L'_{2-2}$ that differ from one by  small
residuals  that vary from point to point.

For the SHFB+LN calculations, the convergence pattern is entirely different.
Here, the ratios of $L_{2-2}/L'_{2-2}$ rapidly converge, but the obtained limit
systematically deviates from one by a small number
$\approx$0.0006. It turns out that this result can be attributed to the
truncation of the quasiparticle spectrum, which is routinely employed in the HFB calculations when dealing with zero-range pairing
interactions. Indeed, such
truncation makes the pairing tensor acquire a small symmetric part, which is
bosonic in character\,\cite{doba05,doba12}. When we switch off the LN
procedure and increase the cutoff energy $E_{\rm cut}$ in such a way that the
variational method is rigorously valid and all quasiparticles are included,
values of $L_{2-2}/L'_{2-2}$ converge perfectly to one as one can assess from  Fig.~\ref{L-i}.

To conclude, small deviations from the exact limit of $L_{2-2}/L'_{2-2}=1$  seen in Fig.~\ref{L-t}
are very well understood, and the KO conditions can be, in fact, met to an
arbitrary precision. Of course, in practical calculations aiming at a
determination of nuclear observables, the precision slightly  below 0.01\% is
perfectly sufficient.

\subsection{Tilted-axis-cranking calculations for triaxial
strongly deformed bands in $^{160}$Yb}

As an illustrative example of our SHF calculations, in this
section we present results obtained for predicted
TSD bands in $^{160}$Yb. A good indicator of unstable PAC solutions
is the appearance of competing PAC minima with similar values of
$\beta_2$ and $|\gamma|$ but opposite values of $\gamma$, as
discussed in Sec.~\ref{intro}. Therefore, before computing the
Routhians as functions of the tilting angle, we first performed
extensive PAC calculations so as to determine deformations of various
minima. Similar to the PAC calculations in $^{158}$Er\,\cite{afan12},
we found that the configurations generally have three typical
deformations, namely, $(Q_t, \gamma) \sim (9\,\mbox{eb},
9^{\circ}$--$14^{\circ})$ (TSD1), $(Q_t, \gamma)
\sim (12.2$--$10.8\,\mbox{eb}, -10^{\circ})$ (TSD2), and $(Q_t, \gamma) \sim (10.0$--$10.5\,\mbox{eb},
13^{\circ})$ (TSD3). Ranges of deformations
indicate shape changes with  rotational frequency.
\begin{table}[htb]
\caption{\label{configs}
The configurations in $^{160}$Yb studied in this work.
Each configuration is described by the numbers of states
occupied in the four parity-signature ($\pi, r$) blocks, in the
convention of Ref.\,\protect\cite{doba04}.
}
\begin{ruledtabular}
\begin{tabular}{cccc}
Label & minimum & Configuration & parity \\
\hline
A & TSD1 & $\nu[23,23,22,22]\otimes\pi[16,18,18,18]$ & + \\
B & TSD1 & $\nu[23,24,21,22]\otimes\pi[16,18,18,18]$ & $-$ \\
C & TSD1 & $\nu[23,24,21,22]\otimes\pi[18,18,17,17]$ & $-$ \\
D & TSD3 & $\nu[23,23,22,22]\otimes\pi[18,18,17,17]$ & + \\
E & TSD3 & $\nu[23,23,22,22]\otimes\pi[17,17,18,18]$ & + \\
\end{tabular}
\end{ruledtabular}
\end{table}

\begin{figure}[htb]
\includegraphics[width=0.35\textwidth]{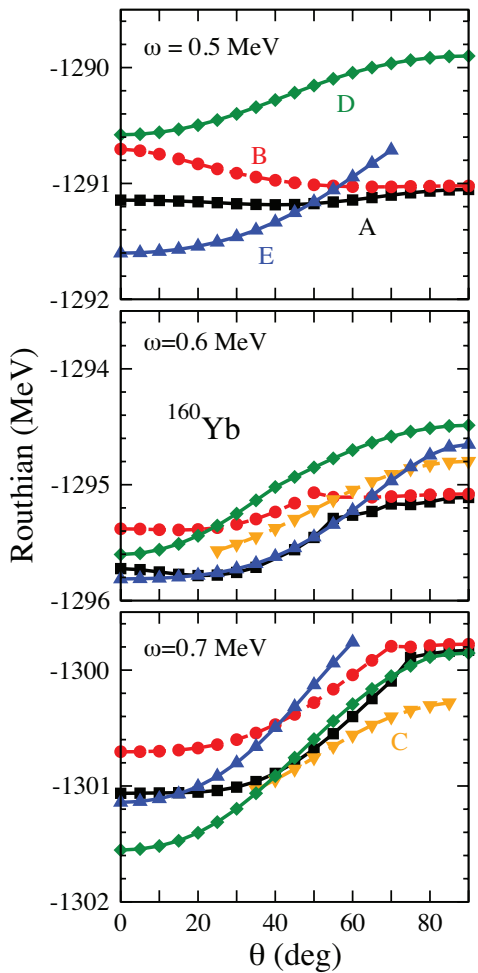}
\caption{(Color online) Total Routhians  in $^{160}$Yb
calculated within the SHF method as functions of the tilting
angle $\theta$  for the five TSD configurations listed in
Table~\protect\ref{configs}. Solid and dashed lines mark configurations
with positive and negative parity, respectively.}
\label{Routhian}
\end{figure}

Figure\,\ref{Routhian} shows total Routhians of five configurations in
$^{160}$Yb calculated in SHFB+LN  as functions of $\theta$. The corresponding
configurations and parities are given in Table~\ref{configs}. At
$\theta=90^{\circ}$, the $Q_{22}$ value changes sign and TSD1 becomes TSD2. It
can be seen that at rotational frequency $\omega = 0.5$\,MeV, the Routhians of
the lowest bands A and B  are very soft against $\theta$. Interestingly, for
the configuration A, a minimum at $\theta\ne$ 0 or 90$^{\circ}$  develops. In
such a situation, one may expect the large-amplitude collective motion of the
rotational axis along $\theta$. The  energies of bands TSD2 rapidly increase
with $\omega$ , and these configurations become saddle points.
For the frequencies considered, the two lowest TSD3 configurations are close
in energy to, or even below, the TSD1 and TSD2 bands.

\section{Conclusions}
\label{conclusions}

In this study we re-examined the Kerman-Onishi conditions for a triaxial
quantum rotation within the self-consistent Kohn-Sham theory. We first derived
the KO equations without invoking the concept of Hamiltonian or wave function
but rather in terms of energy density functional and nucleonic densities. Not
surprisingly, the final form of the KO condition (\ref{eqn3}) is the same as
that originally derived for the HF approximation.

In the next step, we performed numerical tests of the KO condition within the
SHF and SHFB+LN methods. For the first time, these relations have been tested
in a fully self-consistent EDF approach, including pairing. By comparing the
self-consistently obtained Lagrange multiplier $L_{2-2}$ with the value
$L'_{2-2}$ given by Eq.\,(\ref{eqn3}), we demonstrated that in our TAC
calculations, the KO conditions are fulfilled to a very high precision. We
noticed that when differences between $L_{2-2}$ and $L'_{2-2}$ occur, those
are excellent indicators of the variational principle violations due to
practical approximations (such as, e.g., the quasiparticle basis truncation in
HFB).

Finally, we performed 2D TAC calculations for the low-lying TSD bands in
$^{160}$Yb. At lower frequencies, we predict rather $\theta$-soft Routhian
curves, indicative of the large-amplitude collective motion. With increasing
rotational frequency, TSD1 configurations become favored energetically and TSD2
represent unphysical saddle points.
TSD3 configurations become lower in energy than those of TSD1 and TSD2
configurations already at $\omega \approx 0.5$\,MeV, which is much lower than
what has been predicted for $^{158}$Er. A detailed analysis of TSD bands in
$^{160}$Yb will be published elsewhere.

In summary, over thirty years ago,  kinematic conditions for a quantum rotation
of triaxial nuclei were derived within the self-consistent theory.  Now, for
the first time, we have performed EDF calculations that strictly
obey those conditions. Our results are significant for several reasons. First,
they answer a long-standing question in nuclear physics by showing that the
rotations around the axes in a deformed nucleus are not independent of one
another. The advent of computational tools has provided the ample numerical
power to make the numerical calculations of self-consistent TAC  rotations
possible. Second, such calculations are essential for interpreting TSD bands
seen experimentally. Finally, we demonstrated the existence of rotational bands
that are $\theta$-soft, that is, for which the tilting angle cannot be defined.
To understand such structures will require going beyond the single-reference
DFT.

\begin{acknowledgments}
We thank J.A. Sheikh for useful discussions.
This work was supported by the U.S. Department of
Energy (DOE) under Contracts No.\
DE-FG02-96ER40963
(University of Tennessee),  No.\
DE-SC0008499    (NUCLEI SciDAC Collaboration),
Academy of Finland and University of
Jyv\"askyl\"a within the FIDIPRO programme, and
Polish National Science Center under Contract No.\ 2012/07/B/ST2/03907.
\end{acknowledgments}

\bibliographystyle{apsrev4-1}

%

\end{document}